\documentclass[epj]{svjour}
\usepackage{graphics}
\usepackage{psfrag}
\usepackage{epsfig}
\begin{document}
\title{How and when can one identify hadronic molecules in the baryon spectrum}
\author{C. Hanhart
}                     
%
%
\institute{Institut f\"ur Kernphysik (Theorie), Forschungzentrum J\"ulich, D-52425 J\"ulich, Germany}
\date{Received: date / Revised version: date}
%
\abstract{
A method to identify hadronic molecules in the particle spectrum is reviewed
and the conditions for its applicability discussed. Special emphasis is put
on the discussion of molecule candidates in the baryon spectrum.
\PACS{
      {PACS-key}{discribing text of that key}   \and
      {PACS-key}{discribing text of that key}
     } 
} 
\maketitle
\section{Introduction}
\label{intro}

If we want to deduce any information about quark
confinement from the hadron spectrum, it is necessary
to disentangle those states that exist due to
inter--quark (or quark--gluon or gluon--gluon)
 interactions from those that exist due
to hadron--hadron interactions. The latter I will
call hadronic molecules, the former elementary states (sometimes
I will also use the notion quark states, however, it should
become clear that the method outlined only allows
one to quantify the molecular component of a state
and not to draw any conclusion on the composition of
the elementary part, which might be a conventional
meson/baryon or diquark--antidiquark/pentaquark
or contain valence gluons). 
Since also the individual hadrons are made of
quarks and gluons this distinction sounds quite
academic. However, this is not the case.
 In this presentation I
will first discuss under which circumstances
we have a chance to identify hadronic molecules
in a model independent way. After this, the conditions
are checked for various baryons that are candidates
for molecules in the spectrum. 
Then  the example of the $\Lambda (1520)$
is discussed in some detail.

Before going into details some general remarks are
useful. What we call a hadronic molecule\footnote{In Ref.~\cite{jaffe}
the same objects are baptized extraordinary hadrons.}
is an object that exists as the result of non--perturbative
hadron--hadron interactions. For the considerations
below it is not necessary to assume a particular 
mechanism for the hadron--hadron interaction. All
that will be needed is that this interaction is attractive
and sufficiently
strong to form a bound state --- the latter property
will be parameterized by an effective coupling constant.
To understand under which circumstances it is possible
to disentangle hadronic molecules and quark states, 
let us look at the analytic properties of a general
loop digram with either of these degrees of freedom
as possibilities for the essential building block of
a particular state (for simplicity, in this section we  discuss
the case of mesons only --- the generalization of the argument
to baryons is straightforward).
\begin{figure}[t!]
\psfrag{q1}{$q$}
\psfrag{q2}{$\bar q$}
\psfrag{r1}{$= (\mbox{Polynomial in } E)$}
\epsfig{file=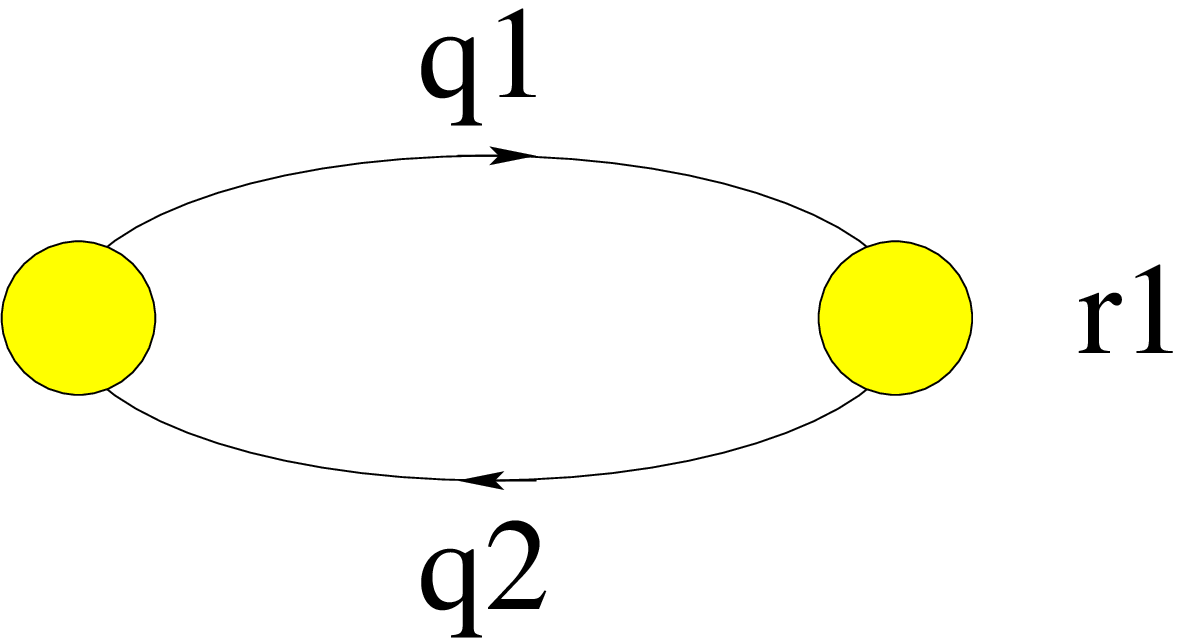,width=0.2\textwidth,angle=0}

\psfrag{q1}{$h_1$}
\psfrag{q2}{$h_2$}
\psfrag{r1}{$= \left\{
\begin{array}{l}
{\ i\mu \sqrt{\phantom{-}E\mu}}+(\mbox{Pol. in } E); \ E>0 \\
{-\mu \sqrt{-E\mu}+(\mbox{Pol. in } E); \ E<0}
\end{array}
\right.$}
\epsfig{file=./loops2.eps,width=0.2\textwidth,angle=0}
\caption{Illustration of the essential difference between
hadron loops (or loops of colour neutral objects) and quark--loops
(or loops of coloured objects): only the former have non--analyticities.}
\label{loops}
\end{figure}
Since quarks can not go on--shell, the corresponding loop integral
needs to lead to a function analytic in the energy.
On the other hand, in addition
to an analytic part,
 the hadronic loop contains  also a non--analytic piece
that originates from the unitarity cut. This piece
is genuine to the two--hadron loop. This situation
is illustrated in Fig. \ref{loops}. Thus, if we can
identify situations where this non--analytic piece
gives the most prominent contribution for molecules,
we should have a chance to disentangle model independently hadronic
from quark loops or, better, hadronic molecules
from elementary states. In the next section we will briefly sketch
how this argument can be made quantitative.

\section{Formalism}
\label{formalism}

The analysis is based on a series of papers by S. Weinberg that
allowed him to deduce that the deuteron can be viewed as a
proton--neutron bound state~\cite{wein}.  The formalism can be
applied, if the binding momentum $\gamma=\sqrt{2\mu\epsilon}$ is much
smaller than any other intrinsic scale of the problem. Here, $\epsilon$
denotes the binding energy with respect to the nearby continuum
channel characterized by its reduced mass $\mu$. Especially,
$\gamma\ll \beta$ is a necessary condition, where $\beta$ denotes the
inverse of the range of forces. Then the low energy scattering
parameters --- or equivalently the effective coupling constant,
$g_{\rm eff}$, of the physical state to the continuum --- provide
a direct measure of the molecular component of the state. It was shown
that (in the normalisation of Ref.~\cite{evidence2})
\begin{eqnarray}
\nonumber
\frac{g^2_{\rm eff}}{4\pi}&=&
16(m_1+m_2)(1-{\cal Z})\sqrt{2\mu \epsilon}
\\ \nonumber
 &\le& 16(m_1+m_2)\sqrt{2\epsilon \mu} \ .
\label{geff}
\end{eqnarray}
where ${\cal Z}$ is a direct measure
of the nature of the state: ${\cal Z}=1$ for
a purely elementary state and ${\cal Z}=0$ for
a molecule. For a short derivation of
the above result see Ref.~\cite{evidence2}.
In Ref.~\cite{evidence} it was argued that 
the same formalism can also be used in the presence
of inelastic channels, as long as the corresponding
thresholds are sufficiently far away and the states
are narrow.
It can be shown~\cite{evidence} that the results described
are equivalent to those of Refs.~\cite{dieanderen}
derived from the theory of scattering off a potential.

When this formalism was applied to the scalar mesons
$a_0(980)$ and $f_0(980)$ it was found from an
analysis of a series of reactions~\cite{ourscalars} that the latter
is indeed predominantly of molecular nature, in line
with the results of Refs.~\cite{mole}, while the
former did not give results that point at an unambiguous
interpretation. This might either point at a prominent
non--molecular contribution to the $a_0$ structure, or
the $a_0$ is not a bound state, but a virtual state.
In Ref.~\cite{Ds} the same ideas were applied to
an analysis of the decays of the $D_s(2317)$ viewed
as a $KD$--molecule. The relation of Eq.~(\ref{geff})
was confirmed in a microscopic model~\cite{osetDs}.
In the same spirit, in Ref.~\cite{X} the data on the hidden--charm
resonance $X(3872)$ was shown to be consistent
with an interpretation as a virtual state in the $D^*\bar D$ system.

In the rest of this paper we discuss the possibilities
of the application of the method to the baryon spectrum,
which was not yet done in detail.

\section{Application to the baryon spectrum}

Recently a sizable number of baryon resonances
was identified as candidates for hadronic molecules. We will
now go through some examples and check for the relevant scales
of the individual problems to see to what extend the method
described in the previous section is applicable.

Probably $the$ classic example for a dynamically generated
baryon state is the $\Lambda (1405)$ --- proposed to originate
from meson--baryon dynamics as early as 1977~\cite{lambda1405_1}.
It is located quite close to the $\Lambda K$ threshold 
($\epsilon \sim 30$ MeV), however,
also the $\pi \Sigma$ threshold is nearby ($\epsilon
\sim 70$ MeV). In addition,
 recently this state was re-investigated within
the chiral unitary approach~\cite{lambda1405_2} with the result that it is 
supposedly composed of two nearby singularities.
This introduces a new, small parameter into the system. For
these two reasons
the formalism of the previous section can not be applied.

The next state of possible dynamical origin is the $\pi N$ resonance
$S_{11}(1535)$~\cite{s111535}. This state is known to couple strongly
to the $\eta N$--channel. However, it is only a resonance in this
channel since it is located above the $\eta N$ threshold --- $m_\eta +
M_N=1487$ MeV --- and it is therefore unlikely to be a dynamical $\eta
N$ state (see also Ref.~\cite{schutz}).  In Ref.~\cite{s111535} the
binding is provided in the strangeness channels $K\Sigma$ and
$K\Lambda$ --- however, those are further away from the pole position
of the $S_{11}(1535)$ --- $m_K+M_\Sigma=1687$ MeV and
$m_K+M_\Lambda=1614$ MeV  --- than the inelastic channel $\eta N$.
Thus  the method
of the previous section can not be applied.

Another state that is a constant problem for quark models
is the Roper resonance $P_{11}(1440)$. In Ref.~\cite{krehl}
it was proposed as a candidate for a dynamical $\sigma N$
state --- where $\sigma$ denotes the very  broad lowest
isoscalar meson resonance ($f_0(600)$ in the particle data
booklet~\cite{PDG}). However, due to the large width of the $\sigma$
not even a well defined threshold can be identified.

However, there are two promising candidates, namely the $\Lambda
(1520)$, proposed to be a $\pi \Sigma(1385)^*$ bound state, and the
$\Delta (1700)$, proposed to be a $\eta \Delta (1232)$ molecule
(although the $K \Sigma(1385)^*$ channel plays a very important
role)~\cite{mesondecuplet1,mesondecuplet2}.  In what follows we will
briefly discuss the former example.

The important parameters to be extracted from the data
are pole positions and residues. Those can not be 
taken directly from the data but theory is needed
for the analytic continuation from the physical axis
into the complex plain where the poles are located.
Thus what is needed for the analysis are not only exclusive
data from various probes in many channels that allow
one to fix the parameters of the theoretical models.
The requirement to these models is a consistency with
analyticity that the analytic continuation can be performed.
The chiral unitary approach meets this requirement
and therefore calculations based on it will be used
for the arguments below.

Let us now focus on the $\Lambda (1520)$.
Please note that the $\Sigma^*(1385)$ is not stable
but has a decay width of $30-40$ MeV (depending on
the charge state~\cite{PDG}). It is not clear how to
take this into account in the relation of Eq. (\ref{geff}).
In the following we will therefore only make a qualitative
comparison with the arguments of Sec.~\ref{formalism}
by adopting that a large effective coupling can be interpreted as an indication
of a large molecular component. 

In Ref.~\cite{mesondecuplet2} the $\Lambda(1520)$ is investigated
within the chiral unitary approach. The channels considered
are $\pi \Sigma^*(1385)$ and $K\Xi^*(1530)$. Thus, with respect 
to the former channel the binding energy is only 5 MeV (if we
take the central value of the mass for the $\Sigma$ and ignore
its width), while the second channel is more than
500 MeV away. Thus the situation seems ideal. The inelastic channels
of relevance are $\bar KN$ and $\pi \Sigma$, both more than 100
MeV away. 

The analysis revealed \cite{mesondecuplet2,witheulogio} that, within
the present approach, indeed the by far largest coupling of the
$\Lambda (1520)$ to a continuum channel is to the $\pi \Sigma^*$
channel. Nevertheless the model is still consistent
with a branching ratio of only 10 \% of the $\Lambda (1520)$ to $\pi
\pi \Lambda$ ($\Lambda \pi$ is with 88\% the most important decay
channel of the $\Sigma^*$), as a result of the limited phase space for
$\pi \Sigma^*$. To test this approach it needs to be applied to
various reactions.  So far a fair description of the data was found
for the reactions $K^-p\to\Lambda\pi\pi$, $\gamma p\to K^+K^-p$,
$\gamma p\to K^+\pi^0\pi^0\Lambda$ and $\pi^- p\to K^0 K^-p$~\cite{lam1520_various}.
In Ref.~\cite{witheulogio} an additional measurement
of  the reactions $pp\to pK^+K^-p$ and $pp\to
pK^+\pi^0\pi^0\Lambda$ was proposed (the calculation
needs additional input, e.g. on baryon--baryon final
state interactions, and was performed on the basis
of the formalism of Ref.~\cite{report}).
In Ref.~\cite{michael}
the radiative decays $\Lambda (1520)\to \gamma \Sigma$ were studied,
however, sizable discrepancies to the data were observed.
Thus the model discussed so far might not be complete yet.

Eventually the model described will be improved sufficiently
that it is consistent with all the available data. Then the
effective coupling constants can be extracted and be compared
to Eq.~(\ref{geff}), once it is understood how to generalize
this equation for unstable decay particles --- then
a solid conclusion can be drawn on the nature of the $\Lambda (1520)$,
however, already now
evidence points at a prominent molecular component.

\section{Summary}

As soon as a physical state is located very close to a continuum
threshold it is possible to extract information on the molecular
component of that state almost directly from the data. 
So far this program was applied in much detail to the 
light scalar mesons, however, also investigations of heavier systems
have started.
The central result of this part was that a large effective coupling
constant of a resonance to a nearby continuum channel shows a large
molecular component of that state. Especially, the effective coupling
constant gets maximal for a pure molecule and its value can be
calculated solely in terms of the binding energy and the masses
of the continuum particles.

In this note for the first time possibilities of a study of the
baryon system were discussed. 
It is argued that the conditions for the applicability of
the above mentioned formalism are not met for most of current
molecule candidates. However,
 the $\Lambda (1520)$
and the $\Delta (1700)$, both already studied within dynamical
models, were identified as promising candidates and the
former was discussed in some detail.
Since it is not known yet how to include a finite width of one
of the continuum particles into the formalism mentioned in the 
previous paragraph, so far no quantitative conclusions regarding
the molecular component of the $\Lambda (1520)$ can be drawn,
however, qualitatively a consistent picture emerges where
it appears as a state generated dynamically in the $\pi \Sigma^*$
channel. 


\end{document}